\documentclass[conference]{IEEEtran}
\IEEEoverridecommandlockouts
\usepackage{cite}
\usepackage{amsmath,amssymb,amsfonts}
\usepackage{algorithmic}
\usepackage{graphicx}
\usepackage{textcomp}
\usepackage{xcolor}
\usepackage{subcaption}
\usepackage{tabularx,ragged2e}

\usepackage[super]{nth}
\def\BibTeX{{\rm B\kern-.05em{\sc i\kern-.025em b}\kern-.08em
    T\kern-.1667em\lower.7ex\hbox{E}\kern-.125emX}}
\begin{document}

\title{Practical Trustworthiness Model for DNN in Dedicated 6G Application
\thanks{This work was partially supported by the DFG Project Nr. 403579441, ”Meteracom: Metrology for parallel THz communication channels.”}
}

\author{
\IEEEauthorblockN{Anouar Nechi$^{1}$, Ahmed Mahmoudi$^{1}$, Christoph Herold$^{2}$, Daniel Widmer$^{1}$, Thomas Kürner$^{2}$, Mladen Berekovic$^{1}$,\\ and Saleh Mulhem$^{1}$}
\IEEEauthorblockA{\textit{$^{1}$Institute of Computer Engineering, University of Lübeck}, Lübeck, Germany \\
\textit{$^{2}$Institute for Communications Technology, Technische Universität Braunschweig}, Braunschweig, Germany \\
$^{1}$\{name.surname\}@uni-luebeck.de, $^{2}$\{surname\}@ifn.ing.tu-bs.de} \\
}
\maketitle

\begin{abstract}
Artificial intelligence (AI) is considered an efficient response to several challenges facing 6G technology. However, AI still suffers from a huge trust issue due to its ambiguous way of making predictions. Therefore, there is a need for a method to evaluate the AI's trustworthiness in practice for future 6G applications. This paper presents a practical model to analyze the trustworthiness of AI in a dedicated 6G application. In particular, we present two customized deep neural networks (DNNs) to solve the automatic modulation recognition (AMR) problem in Terahertz communications-based 6G technology. Then, a specific trustworthiness model and its attributes, namely data robustness, parameter sensitivity, and security covering adversarial examples, are introduced. The evaluation results indicate that the proposed trustworthiness attributes are crucial to evaluate the trustworthiness of DNN for this 6G application.  
\end{abstract}


\section{Introduction}
\label{sec:intro}
The sixth-generation (6G) network technology aims to outperform the current wireless standards by utilizing frequencies above 100 GHz \cite{petrov2020ieee}. Hence, designing efficient communication systems at these frequencies is far more complex than those at lower frequency systems. 6G technology increasingly relies on two main pillars: Terahertz communications (THzCom) and Machine Learning (ML). While 6G has made one further step toward THzCom according to IEEE802.15.3d standard \cite{IEEE_300GHz}, ML is recommended as a novel solution for 6G performance optimization \cite{petrov2020ieee}. In other words, the ultra-wide THz band ranging from 0.1 to 10 THz is foreseen as an excellent candidate for 6G, whereas ML has proved its efficiency in solving technical problems in communication systems \cite{hanna2020combining}. 

Furthermore, ML ambiguously solves several problems in wireless communications. For instance, Deep Neural Networks (DNNs) as a subset of ML have been used in a black-box manner to solve the automatic modulation recognition problem \cite{guo2020open}. Therefore, there is a huge need to understand the risk of deploying such an Artificial Intelligence (AI) algorithm. According to an Independent High-Level Expert Group on AI \cite{ai2019high}, the only possibility to achieve the maximum benefits of AI is to ensure its trustworthiness during all steps of development and use. The concept of trustworthy AI can be perceived as a response to mitigate the risks of deploying AI \cite{thiebes2021trustworthy}. Several works have proposed definitions of system trustworthiness \cite{1335465}, \cite{9869587} or specified definitions of trustworthy AI \cite{ai2019high}, \cite{thiebes2021trustworthy}. However, these definitions are still general and introduce principles more than practical approaches \cite{10.1145/3555803}. In practice, a model of trustworthiness evaluation for a dedicated application in 6G is still missing. To our knowledge, the literature comprises studies investigating AI, especially DNN in communication, such as THzCom-based 6G technology. Nevertheless, the open literature on applying DNN in the 6G domain still needs to catch up to the problem of trustworthiness evaluation.        

\section{Research Methodology \& Background}

Our proposed research methodology is carried out as follows: 
\begin{itemize}
    \item We first define one of the 6G problems. In particular, we choose a THzCom-based automatic modulation recognition (AMR) problem to demonstrate.
    \item We propose two customized DNNs to solve this dedicated problem.
    \item Then, we study trustworthiness attributes that need to be considered for this problem, and we present the so-called trustworthiness model based on these attributes.
    \item Finally, we apply this model as a practical approach to evaluate the trustworthiness of the customized DNNs.
\end{itemize}

The focus of this research methodology is not on developing DNNs to solve the AMR problem but on using the customized DNNs as practical examples to evaluate their trustworthiness in the 6G environment. 
In the following, we introduce THzCom-based AMR as one of the 6G problems, and we review the available DNN-based solutions for such a problem. Then, we investigate the available trustworthiness models for DNNs\@.

\subsection{Deep Learning-based AMR for THz Communication}
In modern communication systems, a transmitter can use a pool of modulation schemes to control data rate and bandwidth usage. While the transmitter adaptively selects the modulation type, the receiver may or may not know the modulation type. This problem is usually perceived as a classification problem, where the receiver aims at recognizing and classifying the modulation. To solve such a problem, modulation information can be supplied in each signal frame, allowing the receiver to identify the modulation type and react accordingly. However, this approach has become more expensive since modern wireless networks are very heterogeneous, and the number of users is increasing significantly. Therefore, such an approach may not be efficient enough in real-world scenarios as it degrades spectrum efficiency due to the additional information in each signal frame \cite{chen2020novel}. 

AMR has been proposed to detect the modulation scheme of received signals without any potential overhead in the network protocol. Ultimately, the signals are demodulated, and the received data is recovered correctly. Further, conventional AMR approaches require a huge amount of computation or experts’ feature extraction experience \cite{xiao2022review}. 
To overcome these issues, deep learning (DL) is considered a powerful tool that can be used for AMR to provide high classification accuracy. DL does not require prior pre-processing or feature extraction, making it more efficient than conventional approaches.
For instance, Convolutional Neural Networks (CNNs) were used in \cite{shi2020deep,gu2019blind,wang2019data} to extract features from raw I/Q data and perform classification. In \cite{guo2020open}, Recurrent Neural Network (RNN)-based AMR has been proposed to extract sequence-correlated features of I/Q signal components and amplitude/phase signal components to recognize modulation schemes. Other works employed RNNs to estimate signal parameters and correct signal distortions like Carrier Frequency Offset (CFO) and multipath fading \cite{hanna2020combining}. The results revealed that the proposed RNN model provides not only good accuracy in signal distortion estimation but also outperforms many DL methods in terms of classification accuracy.

\subsection{AI Trustworthiness}
Several works have investigated the concepts of trustworthiness and dependability to determine their attributes. In \emph{system design}, availability, reliability, safety, integrity, and maintainability are defined as dependability attributes \cite{1335465}. Nevertheless, this definition does not cover all security attributes in which the definition excludes confidentiality. In \cite{9869587}, trustworthiness is defined as a twin of dependability that includes the following attributes: reliability, safety, maintainability, availability, integrity, and confidentiality. This definition considers security as one of the dependability attributes. In \emph{AI-based system design}, the above definitions of trustworthiness do not cover the recent AI requirements. AI is highly data-dependent and needs dedicated attributes for its trustworthiness. Therefore, new attributes of trustworthiness have been introduced, mainly security, robustness, safety, transparency, and fairness \cite{jobin2019global, 10.1145/3555803  }. However, these attributes are general and not specified for a dedicated AI application. To determine the trustworthiness attributes of DNN regarding AMR in THzCom-based 6G technology, the interaction between DNN and its host environment needs to be carefully investigated and described.
\subsection{Paper's Contribution}
As 6G is crossing the primeval stage of its development, it is the right time to consider the trustworthiness of DL deployed in this technology. This paper proposes a trustworthiness model to analyze DNNs designed for recognizing modulation schemes in THzCom-based 6G technology. To the best of our knowledge, this work introduces the first practical approach to evaluating the trustworthiness of DNNs designed for AMR in THzCom-based 6G technology. 

\section{Deep learning-based Automatic Modulation Recognition}
\label{sec:dnnamr}
\subsection{Synthetic THz dataset}
A dataset of transmitted I/Q samples has been used for the AMR task. The THz-dataset contains seven modulation schemes: BPSK, QPSK, 8PSK, QAM16, QAM64, 8APSK, and OOK. Each modulation scheme consists of 26 Signal-to-Noise-Ratio (SNR) levels with 4096 examples per level. The total number of samples in the dataset is \(745,472\). It was generated using the link-level simulation module of the Simulator for Mobile Networks (SiMoNe) \cite{eckhardt2022simone}. The link level simulation module was developed to simulate point-to-point communication links under the influence of realistic propagation effects in accordance with the IEEE802.15.3d standard \cite{IEEE_300GHz}. The simulated transmission was performed using a Root-Raised-Cosine (RRC) transmit pulse and an AWGN channel. The Nyquist Bandwidth is 880 MHz with an oversampling factor of 8, and it has not been subjected to any channel coding technique. All samples have the exact representation to make data processing easier. The THz dataset samples have a 1024 $\times$ 2 shape (I/Q representation).

\subsection{Two DNN Models for Automatic Modulation Classification} 
DNN consists of multiple layers that process input data and generate a set of probabilities (classification). Each layer comprises a set of parameters (weights and biases) used to process the final output in conjunction with the activation function. In the following, two DNNs are adopted using the proposed THz dataset to classify THz modulation schemes. The resulting DNN classifiers have a 32-bit floating point (FP) parameter format.  
\begin{table}[h] 
\caption{CNN network layout}
\centering
\scalebox{0.95}{%
\begin{tabular}{c c} 
\hline
\textbf{Layer} & \textbf{Output dimensions} \\ 
\hline
Input layer & 1024 $\times$ 2 $\times$ 1 \\
Conv2D + BN + ReLU & 1024 $\times$ 2 $\times$ 64 \\
Max Pooling & 512 $\times$ 2 $\times$ 64 \\
Conv2D + BN + ReLU & 512 $\times$ 2 $\times$ 32 \\
Max Pooling & 256 $\times$ 2 $\times$ 32 \\
Conv2D + BN + ReLU & 256 $\times$ 2 $\times$ 16 \\
Max Pooling & 128 $\times$ 2 $\times$ 16 \\
FC/SeLU & 128 \\
FC/SeLU & 128 \\
FC/Softmax & 7 \\
\hline
\end{tabular}}
\label{tab:cnn_layout}
\end{table}

\subsubsection{CNN for AMR}
CNNs have been widely used for computer vision problems \cite{krizhevsky2017imagenet}. A CNN model can learn directly from the raw data without prior expert feature extraction or pre-processing of the raw data. To benefit from this property in AMR, we construct a CNN including three convolutional layers. Each layer is followed by a batch normalization layer, a ReLU activation function, and a MaxPooling layer. We feed the raw I/Q samples of each radio signal into the CNN model. The extracted feature maps are then forwarded to the fully connected region of the network for classification, where we employ the Scaled exponential Linear Unit (SeLU) activation function and an Alpha dropout. The proposed CNN classifier has \(555,287\) parameters, and its layout is shown in Table~\ref{tab:cnn_layout}.   

\begin{figure}[bt]
  \centering
  \includegraphics[scale=0.28]{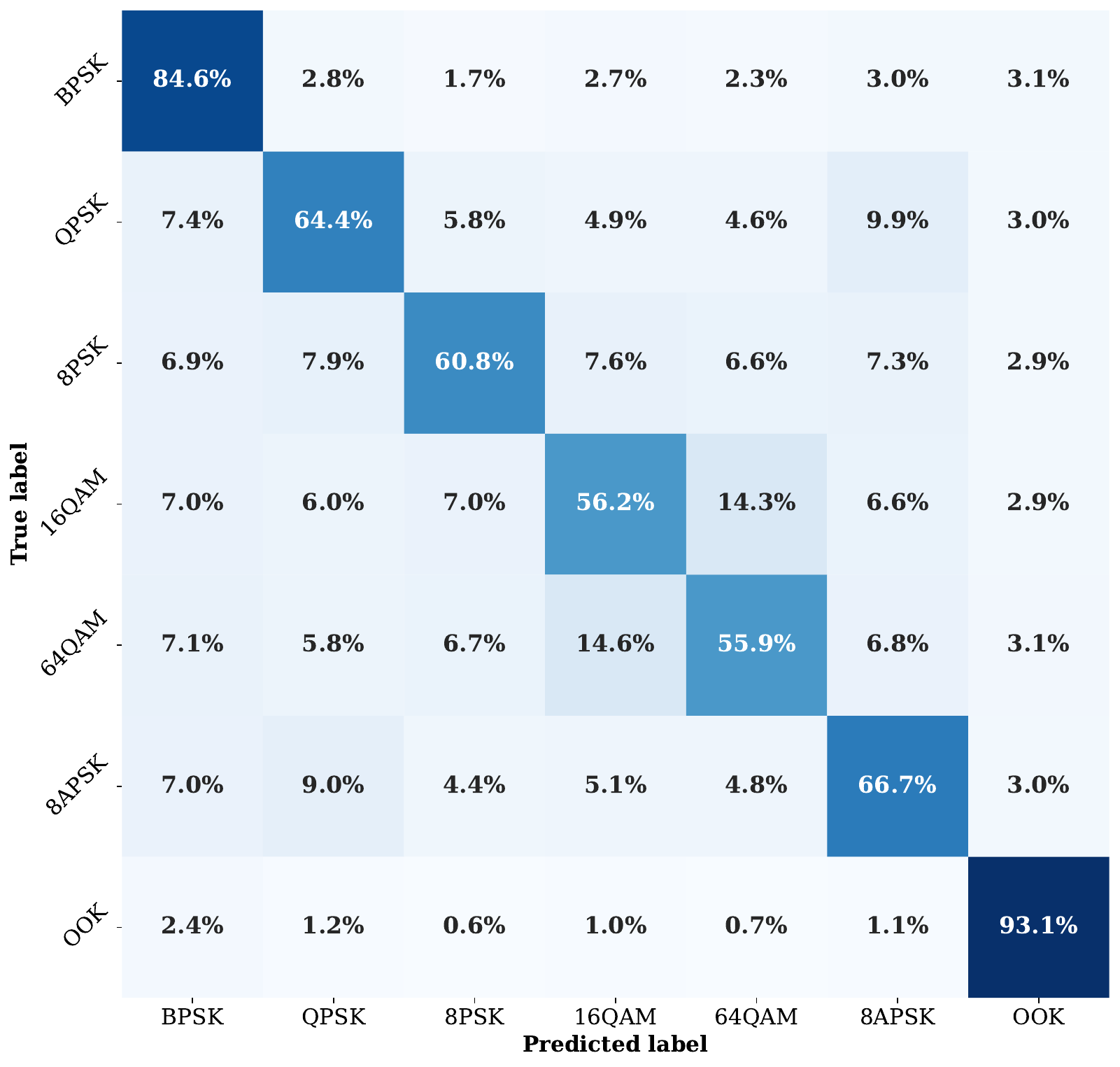}
  \caption{7-Modulation Confusion Matrix for CNN Trained on THz Dataset.}
  \label{fig:cnn_cm}
\end{figure}

The CNN classifier achieves, on average, \(68.8\%\) accuracy across all SNR levels. Fig.~\ref{fig:cnn_cm} shows the confusion matrix of CNN regarding each modulation scheme. CNN classifier imprecisely predicts the higher-order modulation schemes, namely 16QAM and 64QAM, and achieves only \(56.2\%\) and \(55.9\%\) correct predictions, respectively. In contrast, the low-order modulation schemes appear to be the least confused, achieving \(84.6\%\) for BPSK and \(93.1\%\) for OOK.

\subsubsection{ResNet for AMR}
\begin{figure}[bt]
  \centering
  \includegraphics[scale=0.6]{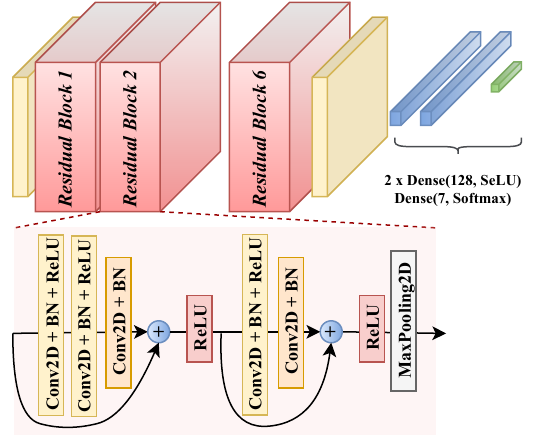}
  \caption{ResNet layout}
  \label{fig:resnet}
\end{figure}

Deep Residual Networks (ResNets) are enhanced versions of CNN. ResNet uses skip connections to process features at multiple scales and depths through the network. Moreover, it is possible to use wider layers, train effectively with fewer epochs, and achieve better results compared to traditional CNN \cite{he2016deep}. We construct a ResNet layout similar to \cite{o2018over} for radio signal classification. Fig.~\ref{fig:resnet} shows the proposed ResNet architecture. It consists of six residual units, each with two skip connections, followed by a fully connected region with the same configuration as the proposed CNN but with only \(159015\) parameters.
\begin{figure}[bt]
  \centering
  \includegraphics[scale=0.28]{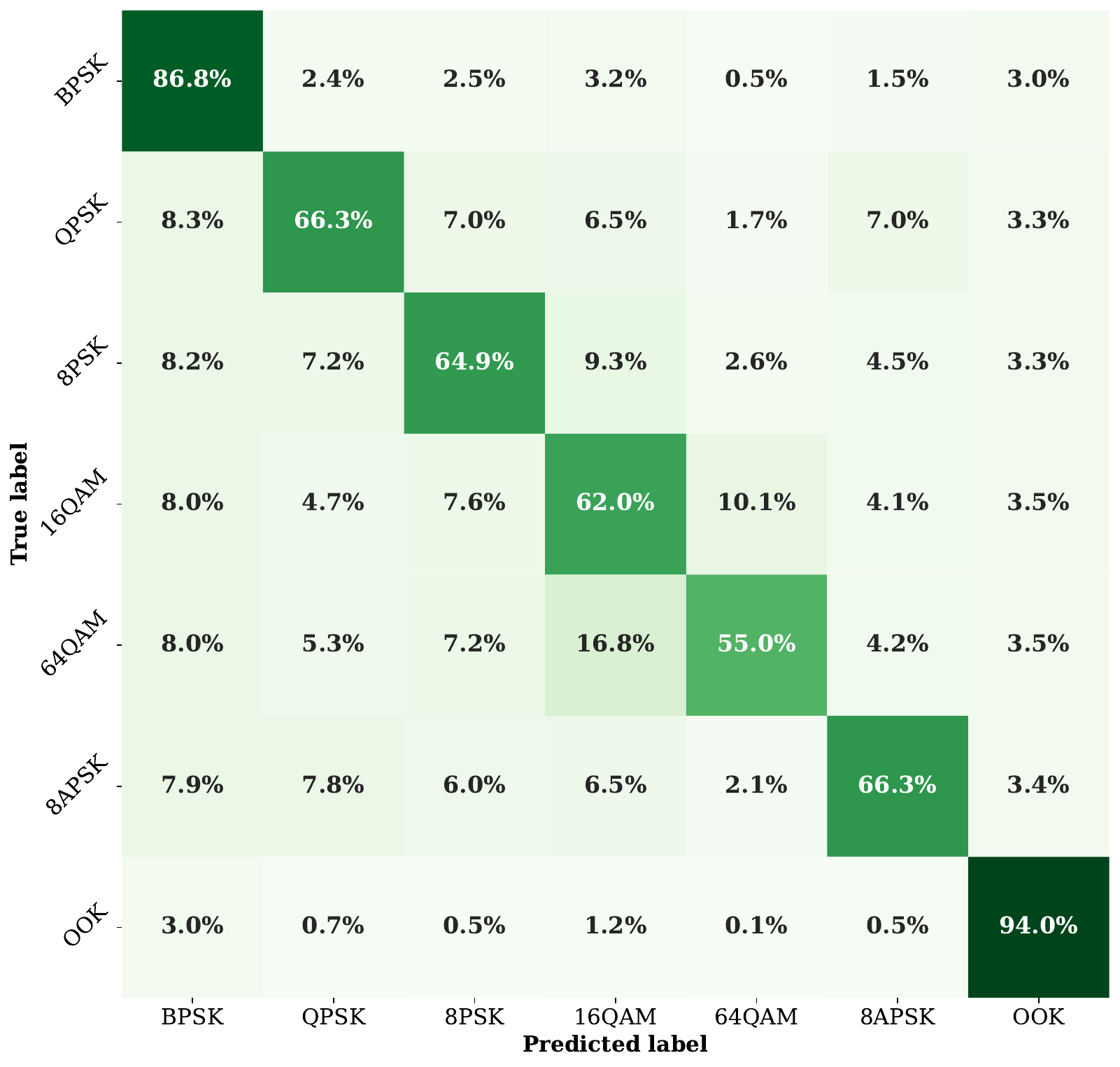}
  \caption{7-Modulation Confusion Matrix for ResNet Trained on THz Dataset.}
  \label{fig:resnet_cm}
\end{figure}

The ResNet classifier achieves \(70.8\%\) accuracy across all SNR levels. ResNet classifier exhibits \(2\%\) higher accuracy than CNN and has fewer parameters. This result emphasizes the effectiveness of ResNets over conventional CNN classifiers. Fig.~\ref{fig:resnet_cm} shows the confusion matrix of ResNet. \(16.8\%\) confusion between 16QAM and 64QAM is noted only. For the remaining modulation schemes, we observe a slight improvement in accuracy.

\section{Trustworthiness: Model \& Attributes}
\label{sec:Trust}
To determine the trustworthiness attributes of DNN regarding AMR in THzCom-based 6G technology, we first formulate DNN as a function of multiple inputs and parameters, and we link this formulation and the trustworthiness attributes as follows.  

The layer \(i\) of a DNN can be seen as an operation \(f_i[p_i](x_{i-1})\), where \(p_i\) represents a set of layer \(i\)'s parameters \(p_i=(W^j_i,b_i)\) including \(j\) weights and one bias, and \(x_{i-1}\) is the output of the previous layer. A composition of these operations defines the DNN classifier \(f_{\textrm{DNN}}\) as, 
\begin{equation}
    f_{\textrm{DNN}}(x_{\textrm{in}}; p)=f_l[p_{l}] \circ \cdots \circ f_2[p_{2}] \circ f_1[p_1](x_{\textrm{in}})
    \label{eq:DNN}
\end{equation}
Where \(x_{\textrm{in}}\) is an input signal, \(p\) is a set of DNN parameters \(p=\{p_1, ..., p_l\}\), and \(l\) is the number of DNN layers. The values of DNN parameters and the model hyperparameters are given during the training phase based on the THz dataset. In the prediction phase, the DNN classifier \(f_{\textrm{DNN}}(x_{\textrm{in}}; p)\) can be perceived as a function of two inputs: The trained parameters \(p\) and the signal \(x_{\textrm{in}}\) like an input variable. 

\begin{figure}[bt]
  \centering
  \includegraphics[width=0.9\linewidth]{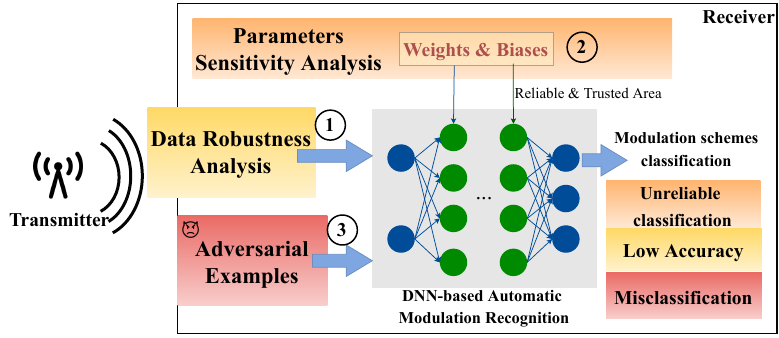}
  \caption{The Proposed Trustworthiness Model of DNN for AMR in 6G: (1) data robustness analysis, (2) parameter sensitivity analysis, and (3) Adversarial examples}
  \label{fig:Trust_f}
\end{figure}

Therefore, the proposed trustworthiness model of such a DNN considers only the input signals \(x_{\textrm{in}}\) and the DNN parameters \(p=\{p_1, ..., p_l\}\). Other building blocks of DNN are considered reliable and trustworthy such as activation functions etc. This model helps to explain how DNN interacts with the THzCom environment and the user. Fig.~\ref{fig:Trust_f} illustrates the three trustworthiness attributes that need to be considered in DNN-based AMR, described as follows: 

\begin{enumerate}
    \item \textbf{Data robustness analysis} helps to understand when the DNN classifiers exhibit low accuracy due to environmental variation. It aims at evaluating such variations and their impact on the quality of DNN classifiers \cite{10.1145/3555803}. Precisely, DNN robustness analysis investigates a noisy environment effect on the input signals \(x_{\textrm{in}}\) and its impact on the DNN classification accuracy. Here, different SNR levels are applied to input signal \(x_{\textrm{in}}\), and then, the drop in the DNN accuracy is observed and estimated. 
    \item \textbf{Parameters sensitivity analysis} provides a deep understanding of DNN's reliability, especially the unreliable classification causes. The reliability can be evaluated by analyzing the sensitivity of the DNN parameters \(p=\{p_1, ..., p_l\}\) for given signals. \textit{Reliability} indicates the DNN classifier should perform with the same accuracy as it's intended without failure (unreliable classification with less than 50\% accuracy). The proposed sensitivity analysis follows a random bit flipping model \cite{Li3126964}.
    \item \textbf{Adversarial examples} indicate the impact of deterministic signal changes on the DNN classifiers introduced by an attacker. Adversarial example attacks can be performed for \textit{Security} evaluation. Here, the attacker chooses and generates inputs \(x_{\textrm{in}}\) to confuse DNN classifiers during the inference phase, resulting in misclassification \cite{SzegedyZSBEGF14}.        
\end{enumerate}

Our trustworthiness model excludes the transparency of DNN from its attributes as AMR doesn't use private data, and it also ignores DNN fairness as the used dataset is balanced. The misclassification leads to selecting an incorrect scheme, and the received signals cannot be demodulated. This event and its consequence are already involved in the reliability attribute. Therefore, DNN safety can be seen as a subset of reliability in our application.

\section{Trustworthiness Analysis of DNN for AMR}
\label{sec:trust}
In this section, we analyze the trustworthiness of the proposed CNN and ResNet by using our trustworthiness model and its attributes.  

\subsection{Data Robustness Analysis}
The impact of environmental variation on the trained DNN model is considered a significant factor of trustworthiness. In other words, the trained DNN model should be aware of the diverse data distribution regarding different environmental scenarios \cite{10.1145/3555803}. In this context, the impact of a noisy environment on DL-based AMR is evaluated. This problem is critical as it affects the data robustness of the DNN model. 

\begin{figure}[b]
  \centering
  \includegraphics[width=0.85\linewidth]{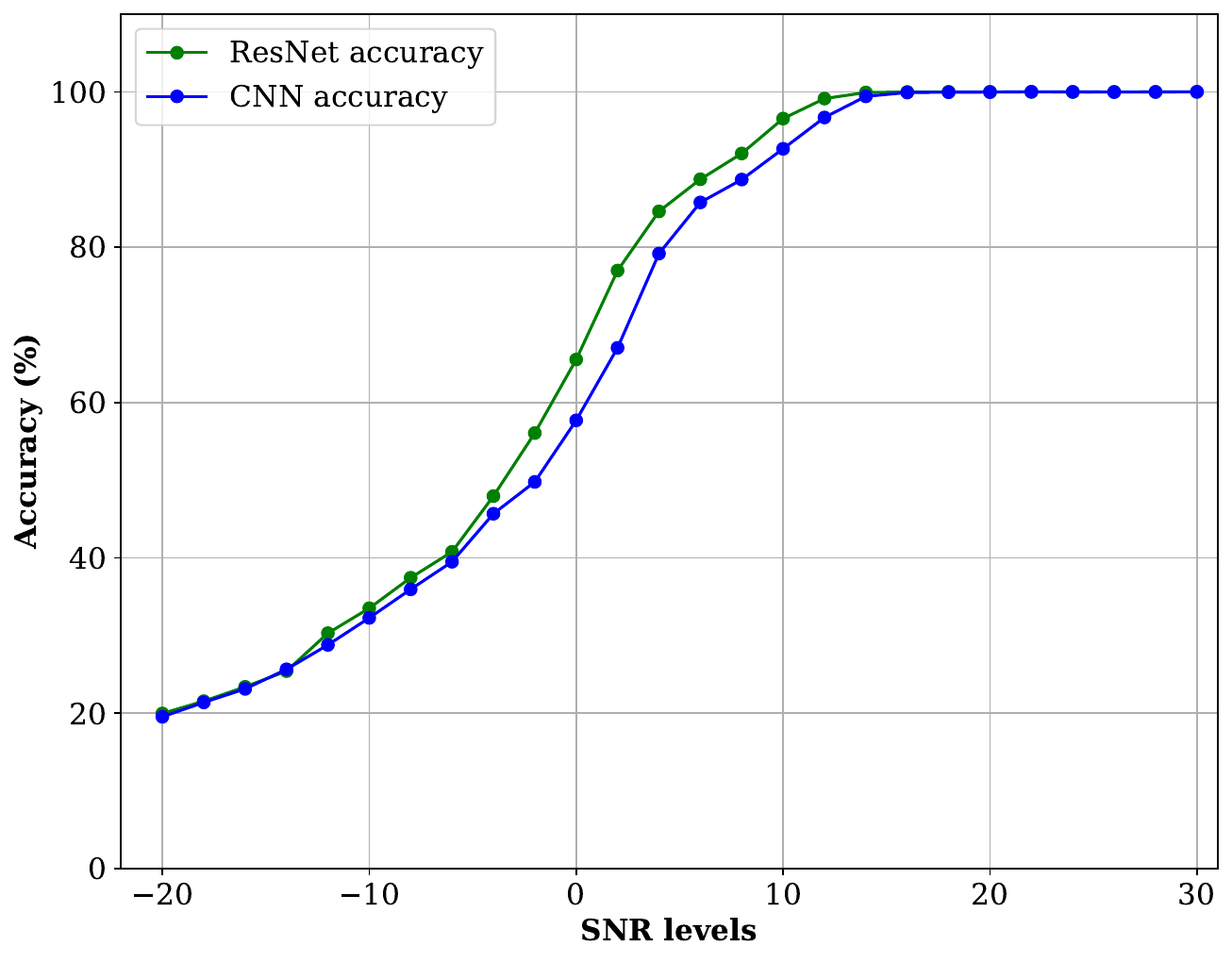}
  \caption{Data Robustness Analysis of both CNN and ResNet on THz Dataset.}
  \label{fig:acc_vs_snr}
\end{figure}

SNR is a crucial metric in any communication system. SNR quantifies the environmental variation by indicating the signal quality concerning a communication channel noise. To analyze the data robustness of DL-based AMR, the following steps are carried out: (1) The dataset is split into a training and testing set with consideration of the various SNR levels to maintain a balanced dataset, (2) DNN models are trained based on the resulting dataset, and (3) the accuracy of the proposed DNN models is evaluated considering the various SNR levels.

Further, we apply the mentioned steps to the proposed CNN and ResNet models. Fig.~\ref{fig:acc_vs_snr} shows that data samples with low SNR ranging from -20 to -4 dB are hard to classify and score a maximum accuracy of 50\%. With such a noise level, the constellation of the received signals is random and does not form meaningful clusters to distinguish between the different modulation schemes. It's worth noting that the model accuracy increases when the SNR increases from -2 dB to 10 dB. The model accuracy attains 99\% as SNR approaches 10 dB. The highest model accuracy is achieved starting from 10 dB.
Moreover, the ResNet model exhibits better accuracy than the CNN model in the SNR interval of -2 dB to 10 dB. The accuracies of both models are correlated out of this interval. As a result, ResNet-based AMR is more robust than CNN-based AMR regarding the noisy channel variation. 

\subsection{Sensitivity Analysis}

AI sensitivity analysis determines vulnerable bits that significantly decrease the classification accuracy when flipped. Sensitivity analysis relies on a bit-flipping model of AI parameters. Sensitivity analysis aims to provide a deep understanding of AI's behavior and gives some hints towards explaining AI's decision-making.

To conduct the sensitivity analysis of CNN and ResNet classifiers, a single-bit flip is randomly introduced to the DNN's parameters \cite{Li3126964}. Both the bit position and the targeted parameter are uniformly distributed. First, we randomly inject single-bit faults 1000 times at different bit positions and parameter locations of each layer of the CNN classifier. In the case of the ResNet, we randomly inject single-bit faults in the residual block, convolution, and dense layers. Nevertheless, the injected faults in the convolution and dense layers are performed similarly to the CNN. However, the faults in the residual block are randomly injected at different bit positions, parameter locations, and layers. The above fault injection experiments are conducted during inference.

\begin{figure}[bt]
  \centering
  \includegraphics[width=0.95\linewidth]{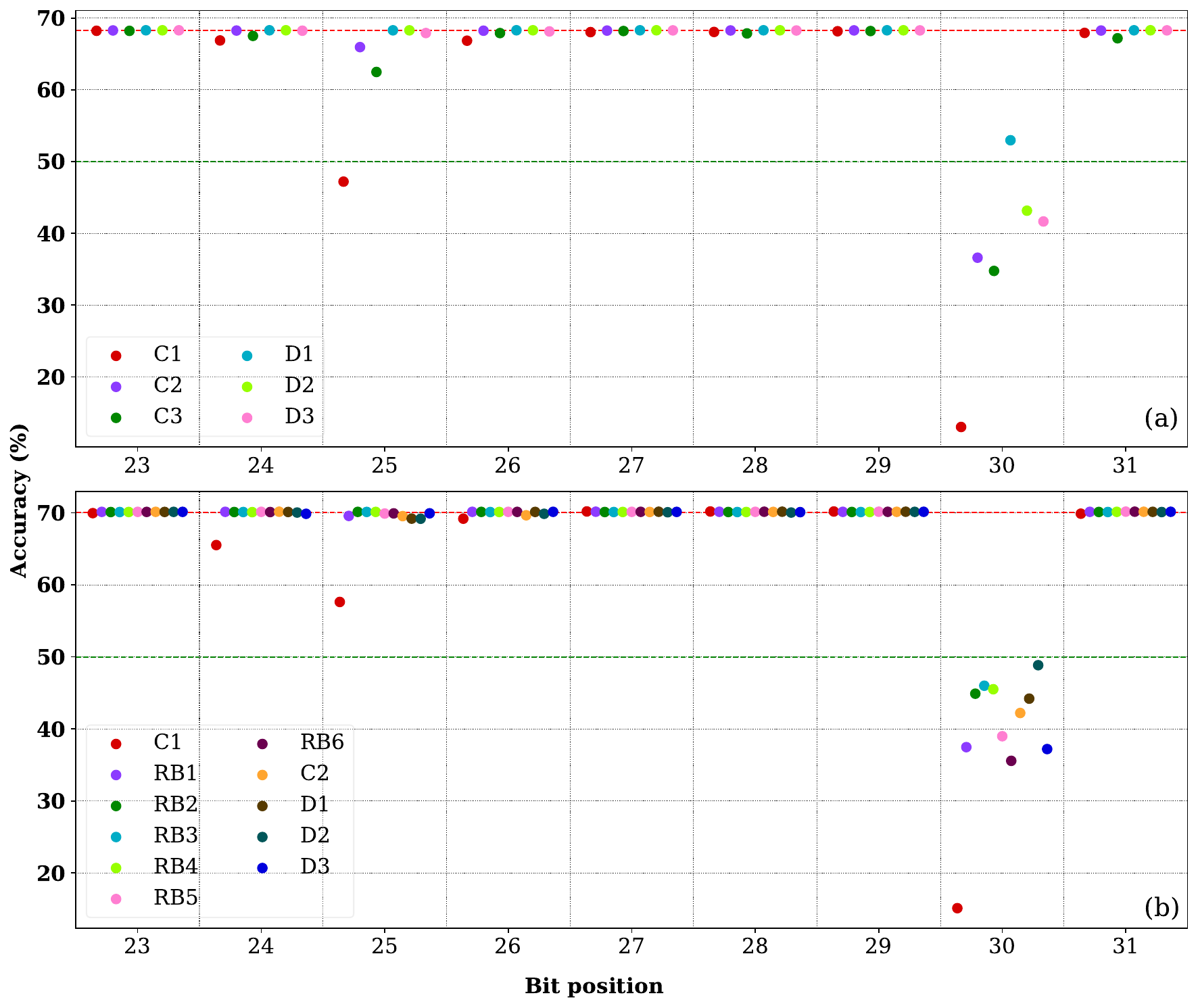}
  \caption{Weight Sensitivity of (a) CNN, and (b) ResNet Classifiers }
  \label{fig:fi}
\end{figure}

The single-bit faults injected in 32-bit FP parameters indicate that the exponent (from 23 to 30-bit position) is more sensitive than the mantissa (from 0 to 22-bit position). This emphasizes the well-known results in public literature. 
To better understand the exponent sensitivity, we divide the vulnerable exponent bits into two categories: the first category includes the vulnerable bits resulting in misclassification (unreliable classifier with accuracy lower than 50\%), and the second covers the vulnerable bits resulting in accuracy degradation.

Fig. \ref{fig:fi}-a illustrates the impact of single-bit faults on the CNN classifier regarding convolution layers {C1, C2, C3} and dense layers {D1, D2, D3}. The unreliable classification is observed at 25 in C1 and 30 in {C1, C2, C3, D2, D3}. However, the flipping of bit 30 in D1 results only in accuracy degradation. It should be noted that the faults injected in the remaining layers show insignificant accuracy degradation. Fig. \ref{fig:fi}-b shows the impact of bit flipping on all layers of the ResNet classifier. The 30th bit (i.e.\ bit 31) is more sensitive than the others as it causes unreliable classification across all layers. The remaining vulnerable bits cause only an accuracy drop.

As a result, flipping the vulnerable 30th-bit causes the misclassification of both classifiers. Other vulnerable bits show only accuracy degradation.

\subsection{Security Analysis}

In \cite{SzegedyZSBEGF14}, several neural network models exhibit vulnerabilities to adversarial examples, where the attacker generates some inputs that lead to misclassification. These inputs are slightly different from the original inputs that are classified correctly, yet they are likely to cause such misclassification. The adversarial examples mainly occur due to some \textit{``linear behavior in high-dimensional spaces''}~\cite{GoodfellowSS14}. This observation introduces many efficient adversarial example attacks such as Fast Gradient Method~\cite{GoodfellowSS14}, Projected Gradient Descend~\cite{kurakin2018adversarial18}. 

\begin{table}[h]  
\caption{Results of Adversarial Examples Attacks}
\centering
\scalebox{0.95}{%
\begin{tabular}{ccc}
\hline
\textbf{Attack}                         & $\textbf{AER}_{\textbf{CNN}}$     & $\textbf{AER}_{\textbf{ResNet}}$\\
\hline
FGM~\cite{GoodfellowSS14}               & $18.83\%$                         & $22.12\%$\\
PGD~\cite{kurakin2018adversarial18}     & $6.11\%$                          & $5.60\%$\\
NewtonFool~\cite{10.1145/3134600.3134635} & $26.84\%$                       & $28.50\%$\\
DeepFool~\cite{Moosavi-Dezfooli16}      & $26.07\%$                         & $15.43\%$\\
HopSkipJump~\cite{9152788}              & $14.92\%$                         & $8.04\%$\\
Zoo~\cite{10.1145/3128572.3140448}      & $26.57\%$                         & $19.25\%$\\
C\&W~\cite{7958570} \(L_2\)             & $53.9\%$                          & $59.25\%$\\
C\&W~\cite{7958570} \(L_\infty\)        & $21.57\%$                         & $14.57\%$\\
\hline
\end{tabular}}
\label{tab:Attacks}
\end{table}

To analyze the impact of adversarial examples, we launch eight attacks to generate adversarial examples against the investigated CNN and ResNet classifiers using the Adversarial Robustness Toolbox (v1.2.0)~\cite{art2018}. We set up each attack using the predefined sets of attack parameters. Then, we perform the same attacks on both classifiers. Table~\ref{tab:Attacks} shows the attack results of Fast Gradient Method (FGM)~\cite{GoodfellowSS14}, Projected Gradient Descend (PGD)~\cite{kurakin2018adversarial18}, NewtonFool~\cite{10.1145/3134600.3134635}, DeepFool~\cite{Moosavi-Dezfooli16}, HopSkipJump~\cite{9152788}, Zeroth Order Optimization (Zoo)~\cite{10.1145/3128572.3140448}, and Carlini \& Wagner methods (C\&W)~\cite{7958570} over \(L_2\), and \(L_{\infty}\). The adversarial example resistance (AER) quantifies the model's correct classifications despite adversarial example generation. For instance, both classifiers exhibit comparably high AER against C\&W over \(L_2\), while the generated adversarial examples by PGD fundamentally devastate both. Generally, both models lead to different AER w.r.t.\ the respective attacks. Thus, the decision for the more-resistant model against adversarial examples will remain dependent on the chosen attacks.

\subsection{Trustworthiness Evaluation}

According to the above analysis, the trustworthiness level of the investigated classifiers can be evaluated as follows. First, when the SNR ranges from 0 to 30 dB, both classifiers are robust against environmental variations. Here, ResNet shows greater robustness than the CNN classifier. Second, the sensitivity analysis of the classifier’s parameters indicates that flipping the vulnerable 30th-bit results in unreliable classifications. Finally, both classifiers show different levels of resistance against selected adversarial example attacks without yielding a clear verdict for either DNN. Future work may identify suitable attacks to provide the required metrics.
\section{Conclusion}
\label{sec:conclusion}



In this paper, we introduced a methodology to build a practical trustworthiness model of Deep Neuronal Networks (DNN) dedicated to one of the 6G applications. The need for such a model is significant as 6G technology requires higher levels of reliability and security compared to prior generations.

Particularly, we constructed two DNN classifiers addressing the automatic modulation recognition (AMR) problem in a THzCom-based 6G environment. Then, we applied our trustworthiness model to analyze the classifiers w.r.t. attributes chosen to meet this environment: Robustness, DNN parameter reliability, and DNN adversarial example resistance. Based on our experiments results, we conclude our trustworthiness model a suitable approach to analyze the trustworthiness of the used DNN for AMR in THzCom-based 6G technology.

\bibliographystyle{IEEEtran}
\bibliography{ref.bib}

\begin{thebibliography}{10}
\providecommand{\url}[1]{#1}
\csname url@samestyle\endcsname
\providecommand{\newblock}{\relax}
\providecommand{\bibinfo}[2]{#2}
\providecommand{\BIBentrySTDinterwordspacing}{\spaceskip=0pt\relax}
\providecommand{\BIBentryALTinterwordstretchfactor}{4}
\providecommand{\BIBentryALTinterwordspacing}{\spaceskip=\fontdimen2\font plus
\BIBentryALTinterwordstretchfactor\fontdimen3\font minus
  \fontdimen4\font\relax}
\providecommand{\BIBforeignlanguage}[2]{{%
\expandafter\ifx\csname l@#1\endcsname\relax
\typeout{** WARNING: IEEEtran.bst: No hyphenation pattern has been}%
\typeout{** loaded for the language `#1'. Using the pattern for}%
\typeout{** the default language instead.}%
\else
\language=\csname l@#1\endcsname
\fi
#2}}
\providecommand{\BIBdecl}{\relax}
\BIBdecl

\bibitem{petrov2020ieee}
V.~Petrov, T.~Kurner, and I.~Hosako, ``Ieee 802.15. 3d: First standardization
  efforts for sub-terahertz band communications toward 6g,'' \emph{IEEE
  Communications Magazine}, vol.~58, no.~11, pp. 28--33, 2020.

\bibitem{IEEE_300GHz}
``Ieee standard for high data rate wireless multi-media networks--amendment 2:
  100 gb/s wireless switched point-to-point physical layer,'' \emph{IEEE Std
  802.15.3d-2017 (Amendment to IEEE Std 802.15.3-2016 as amended by IEEE Std
  802.15.3e-2017)}, pp. 1--55, 2017.

\bibitem{hanna2020combining}
S.~Hanna, C.~Dick, and D.~Cabric, ``Combining deep learning and linear
  processing for modulation classification and symbol decoding,'' in
  \emph{GLOBECOM 2020-2020 IEEE Global Communications Conference}.\hskip 1em
  plus 0.5em minus 0.4em\relax IEEE, 2020, pp. 1--7.

\bibitem{guo2020open}
Y.~Guo, H.~Jiang, J.~Wu, and J.~Zhou, ``Open set modulation recognition based
  on dual-channel lstm model,'' \emph{arXiv preprint arXiv:2002.12037}, 2020.

\bibitem{ai2019high}
A.~HLEG, ``High-level expert group on artificial intelligence, ethics
  guidelines for trustworthy ai,'' AI HLEG, Tech. Rep., 2019.

\bibitem{thiebes2021trustworthy}
S.~Thiebes, S.~Lins, and A.~Sunyaev, ``Trustworthy artificial intelligence,''
  \emph{Electronic Markets}, vol.~31, no.~2, pp. 447--464, 2021.

\bibitem{1335465}
A.~Avizienis, J.-C. Laprie, B.~Randell, and C.~Landwehr, ``Basic concepts and
  taxonomy of dependable and secure computing,'' \emph{IEEE Transactions on
  Dependable and Secure Computing}, vol.~1, no.~1, pp. 11--33, 2004.

\bibitem{9869587}
B.~Bauer, M.~Ayache, S.~Mulhem, M.~Nitzan, J.~Athavale, R.~Buchty, and
  M.~Berekovic, ``On the dependability lifecycle of electrical/electronic
  product development: The dual-cone v-model,'' \emph{Computer}, vol.~55,
  no.~9, pp. 99--106, 2022.

\bibitem{10.1145/3555803}
\BIBentryALTinterwordspacing
B.~Li, P.~Qi, B.~Liu, S.~Di, J.~Liu, J.~Pei, J.~Yi, and B.~Zhou, ``Trustworthy
  ai: From principles to practices,'' \emph{ACM Comput. Surv.}, aug 2022, just
  Accepted. [Online]. Available: \url{https://doi.org/10.1145/3555803}
\BIBentrySTDinterwordspacing

\bibitem{chen2020novel}
S.~Chen, Y.~Zhang, Z.~He, J.~Nie, and W.~Zhang, ``A novel attention cooperative
  framework for automatic modulation recognition,'' \emph{IEEE Access}, vol.~8,
  pp. 15\,673--15\,686, 2020.

\bibitem{xiao2022review}
W.~Xiao, Z.~Luo, and Q.~Hu, ``A review of research on signal modulation
  recognition based on deep learning,'' \emph{Electronics}, vol.~11, no.~17, p.
  2764, 2022.

\bibitem{shi2020deep}
J.~Shi, S.~Hong, C.~Cai, Y.~Wang, H.~Huang, and G.~Gui, ``Deep learning-based
  automatic modulation recognition method in the presence of phase offset,''
  \emph{IEEE Access}, vol.~8, pp. 42\,841--42\,847, 2020.

\bibitem{gu2019blind}
H.~Gu, Y.~Wang, S.~Hong, and G.~Gui, ``Blind channel identification aided
  generalized automatic modulation recognition based on deep learning,''
  \emph{IEEE Access}, vol.~7, pp. 110\,722--110\,729, 2019.

\bibitem{wang2019data}
Y.~Wang, M.~Liu, J.~Yang, and G.~Gui, ``Data-driven deep learning for automatic
  modulation recognition in cognitive radios,'' \emph{IEEE Transactions on
  Vehicular Technology}, vol.~68, no.~4, pp. 4074--4077, 2019.

\bibitem{jobin2019global}
A.~Jobin, M.~Ienca, and E.~Vayena, ``The global landscape of ai ethics
  guidelines,'' \emph{Nature Machine Intelligence}, vol.~1, no.~9, pp.
  389--399, 2019.

\bibitem{eckhardt2022simone}
\BIBentryALTinterwordspacing
J.~M. Eckhardt, C.~Herold, B.~K. Jung, N.~Dreyer, and T.~Kürner, ``Modular
  link level simulator for the physical layer of beyond 5g wireless
  communication systems,'' \emph{Radio Science}, vol.~57, no.~2, p.
  e2021RS007395, 2022, e2021RS007395 2021RS007395. [Online]. Available:
  \url{https://agupubs.onlinelibrary.wiley.com/doi/abs/10.1029/2021RS007395}
\BIBentrySTDinterwordspacing

\bibitem{krizhevsky2017imagenet}
A.~Krizhevsky, I.~Sutskever, and G.~E. Hinton, ``Imagenet classification with
  deep convolutional neural networks,'' \emph{Communications of the ACM},
  vol.~60, no.~6, pp. 84--90, 2017.

\bibitem{he2016deep}
K.~He, X.~Zhang, S.~Ren, and J.~Sun, ``Deep residual learning for image
  recognition,'' in \emph{Proceedings of the IEEE conference on computer vision
  and pattern recognition}, 2016, pp. 770--778.

\bibitem{o2018over}
T.~J. O’Shea, T.~Roy, and T.~C. Clancy, ``Over-the-air deep learning based
  radio signal classification,'' \emph{IEEE Journal of Selected Topics in
  Signal Processing}, vol.~12, no.~1, pp. 168--179, 2018.

\bibitem{Li3126964}
\BIBentryALTinterwordspacing
G.~Li, S.~K.~S. Hari, M.~Sullivan, T.~Tsai, K.~Pattabiraman, J.~Emer, and S.~W.
  Keckler, ``Understanding error propagation in deep learning neural network
  (dnn) accelerators and applications,'' ser. SC '17.\hskip 1em plus 0.5em
  minus 0.4em\relax New York, NY, USA: Association for Computing Machinery,
  2017. [Online]. Available: \url{https://doi.org/10.1145/3126908.3126964}
\BIBentrySTDinterwordspacing

\bibitem{SzegedyZSBEGF14}
\BIBentryALTinterwordspacing
C.~Szegedy, W.~Zaremba, I.~Sutskever, J.~Bruna, D.~Erhan, I.~J. Goodfellow, and
  R.~Fergus, ``Intriguing properties of neural networks,'' in \emph{2nd
  International Conference on Learning Representations, {ICLR} 2014, Banff, AB,
  Canada, April 14-16, 2014, Conference Track Proceedings}, Y.~Bengio and
  Y.~LeCun, Eds., 2014. [Online]. Available:
  \url{http://arxiv.org/abs/1312.6199}
\BIBentrySTDinterwordspacing

\bibitem{GoodfellowSS14}
\BIBentryALTinterwordspacing
I.~J. Goodfellow, J.~Shlens, and C.~Szegedy, ``Explaining and harnessing
  adversarial examples,'' in \emph{3rd International Conference on Learning
  Representations, {ICLR} 2015, San Diego, CA, USA, May 7-9, 2015, Conference
  Track Proceedings}, Y.~Bengio and Y.~LeCun, Eds., 2015. [Online]. Available:
  \url{http://arxiv.org/abs/1412.6572}
\BIBentrySTDinterwordspacing

\bibitem{kurakin2018adversarial18}
A.~Kurakin, I.~J. Goodfellow, and S.~Bengio, ``Adversarial examples in the
  physical world,'' in \emph{Artificial intelligence safety and
  security}.\hskip 1em plus 0.5em minus 0.4em\relax Chapman and Hall/CRC, 2018,
  pp. 99--112.

\bibitem{10.1145/3134600.3134635}
\BIBentryALTinterwordspacing
U.~Jang, X.~Wu, and S.~Jha, ``Objective metrics and gradient descent algorithms
  for adversarial examples in machine learning,'' in \emph{Proceedings of the
  33rd Annual Computer Security Applications Conference}, ser. ACSAC '17.\hskip
  1em plus 0.5em minus 0.4em\relax New York, NY, USA: Association for Computing
  Machinery, 2017, p. 262–277. [Online]. Available:
  \url{https://doi.org/10.1145/3134600.3134635}
\BIBentrySTDinterwordspacing

\bibitem{Moosavi-Dezfooli16}
\BIBentryALTinterwordspacing
S.~Moosavi{-}Dezfooli, A.~Fawzi, and P.~Frossard, ``Deepfool: {A} simple and
  accurate method to fool deep neural networks,'' in \emph{2016 {IEEE}
  Conference on Computer Vision and Pattern Recognition, {CVPR} 2016, Las
  Vegas, NV, USA, June 27-30, 2016}.\hskip 1em plus 0.5em minus 0.4em\relax
  {IEEE} Computer Society, 2016, pp. 2574--2582. [Online]. Available:
  \url{https://doi.org/10.1109/CVPR.2016.282}
\BIBentrySTDinterwordspacing

\bibitem{9152788}
J.~Chen, M.~I. Jordan, and M.~J. Wainwright, ``Hopskipjumpattack: A
  query-efficient decision-based attack,'' in \emph{2020 IEEE Symposium on
  Security and Privacy (SP)}, 2020, pp. 1277--1294.

\bibitem{10.1145/3128572.3140448}
\BIBentryALTinterwordspacing
P.-Y. Chen, H.~Zhang, Y.~Sharma, J.~Yi, and C.-J. Hsieh, ``Zoo: Zeroth order
  optimization based black-box attacks to deep neural networks without training
  substitute models,'' in \emph{Proceedings of the 10th ACM Workshop on
  Artificial Intelligence and Security}, ser. AISec '17.\hskip 1em plus 0.5em
  minus 0.4em\relax New York, NY, USA: Association for Computing Machinery,
  2017, p. 15–26. [Online]. Available:
  \url{https://doi.org/10.1145/3128572.3140448}
\BIBentrySTDinterwordspacing

\bibitem{7958570}
\BIBentryALTinterwordspacing
N.~Carlini and D.~Wagner, ``Towards evaluating the robustness of neural
  networks,'' in \emph{2017 IEEE Symposium on Security and Privacy (SP)}.\hskip
  1em plus 0.5em minus 0.4em\relax Los Alamitos, CA, USA: IEEE Computer
  Society, may 2017, pp. 39--57. [Online]. Available:
  \url{https://doi.ieeecomputersociety.org/10.1109/SP.2017.49}
\BIBentrySTDinterwordspacing

\bibitem{art2018}
\BIBentryALTinterwordspacing
M.-I. Nicolae, M.~Sinn, M.~N. Tran, B.~Buesser, A.~Rawat, M.~Wistuba,
  V.~Zantedeschi, N.~Baracaldo, B.~Chen, H.~Ludwig, I.~Molloy, and B.~Edwards,
  ``Adversarial robustness toolbox v1.2.0,'' \emph{CoRR}, vol. 1807.01069,
  2018. [Online]. Available: \url{https://arxiv.org/pdf/1807.01069}
\BIBentrySTDinterwordspacing

\end{thebibliography}

\end{document}